\documentclass[reqno,aps,prd,onecolumn,showpacs,showkeys,preprintnumbers,amsmath,amssymb,amsfonts]{revtex4}

\usepackage{graphics}
\usepackage[dvips]{graphicx}
\usepackage{dcolumn}
\usepackage{bm}
\usepackage[bookmarksnumbered]{hyperref}
\usepackage{color}

\numberwithin{equation}{section}

\begin{document}

\preprint{Quk5.1a}

\pdfbookmark{Quarkonium and hydrogen spectra with spin dependent relativistic wave equation}{tit}
\title{Quarkonium and hydrogen spectra with spin\\ dependent relativistic wave equation}


\author{Vikram H. Zaveri}
\email{zaverivik@hotmail.com}
\affiliation{B-4/6, Avanti Apt., Harbanslal Marg, Sion, Mumbai 400022 INDIA}


Journal: PRAMANA -- J. Phys. v.75(4), (2010), 579--598.\ \ \ 
DOI: \href{http://dx.doi.org/10.1007/s12043-010-0140-6}{10.1007/s12043-010-0140-6}

\date{March 22, 2010}

\begin{abstract}
A non-linear non-perturbative relativistic atomic theory introduces spin in the  dynamics of particle motion. The resulting energy levels of Hydrogen atom are exactly same as the Dirac theory. The theory accounts for the energy due to spin-orbit interaction and for the additional potential energy due to spin and spin-orbit coupling. Spin angular momentum operator is integrated into the equation of motion. This requires modification to classical Laplacian operator. Consequently the Dirac matrices and the k operator of Dirac's theory are dispensed with. The theory points out that the curvature of the orbit draws on certain amount of kinetic and potential energies affecting the momentum of electron and the spin-orbit interaction energy constitutes a part of this energy. The theory is developed for spin 1/2 bound state single electron in Coulomb potential and then further extended to quarkonium physics by introducing the linear confining potential. The unique feature of this quarkonium model is that the radial distance can be exactly determined and does not have a statistical interpretation. The established radial distance is then used to determine the wave function. The observed energy levels are used as the input parameters and the radial distance and the string tension are predicted. This ensures 100\% conformance to all observed energy levels for the heavy quarkonium.
\end{abstract}


\maketitle
\section{Introduction}
\hspace*{5 mm} It has been more than three quarters of a century since the first publication of Schr$\ddot{o}$dinger theory \cite{7} and apart from Dirac's theory \cite{3,8,16} no other satisfactory solution to the fine structure problem of hydrogen has been found. Specifically, attempts at introducing spin in unperturbed relativistic Schr$\ddot{o}$dinger formalism were unsuccessful. In the earlier work \cite{61}, we argued that the classical Laplacian operator of the Schr$\ddot{o}$dinger theory \cite{7} does not provide sufficient description for the total momentum of the particle and needs to be revised to include the effect of spin. In Dirac's theory \cite{3,8,16} we provide correction to del operator from outside by introducing Dirac matrices, k operator and by making a proper choice in selecting the radial momentum operator \cite{16,14,15,17,18}. In this theory, the changes have been introduced from within the Laplacian. In other words the spin is included into the dynamics of motion. The curvature of the orbit is intorduced by proposing a novel form of separation constant for the bound state radial wave equation, which reduces to standard value $E$ for a free particle. The resulting energy levels of Hydrogen atom are exactly same as that of the Dirac theory.\\

In this work the theory is further extended to include the linear confining potential with significant results in explaining the quarkonium spectra. Variety of potential models \cite{94,63,64,65,66,67,70,71,72,73,76,77,78,80,95,82,83,86,88,89,90,91,92,93} have been presented to explain the quarkonium spectra. Most models are based on non-relativistic or semi-relativistic Schr$\ddot{o}$dinger wave equation. Some use Dirac's wave equation, but all have one thing in common. The spin is introduced with the help of perturbative QCD. A phenomenological model presented here introduces spin directly into the Laplacian and takes into account other relativistic effects right into the foundation of the theory and not as a perturbative attachment. The model explains hydrogen and heavy quarkonium with the same form of wave function. The unique feature of this quarkonium model is that the radial distance can be exactly determined and does not have a statistical interpretation like the hydrogen model, even though both models have a common theoretical plateform. All other models listed above predict $rms$ radial distance which has statistical interpretation like the hydrogen atom.

Sec.1 introduces the scope of this paper. Sec.2 introduces the spin and the curvature into the new relativistic wave equation. Sec.3 presents application of the new theory to the hydrogen atom. Sec.4 provides extension of the theory to the  quarkonium physics. Sec.5 presents application of the new theory to the quarkonium spectra. Sec.6 presents concluding remarks.

\section{Spin dependent relativistic wave equation}

Using the energy momentum invariant of a particle, we write the wave equation in the form
\begin{align}\label{9.1}
E\psi=\pm[(c\:\boldsymbol{p})^2 + (m_o c^2)^2]^{1/2}\psi+V\psi,
\end{align}
where $\pm$ sign indicates positive and negative energies of Dirac theory.
Here $E$ represents the total orbital energy of the particle consisting of kinetic energy $E'$, rest energy $m_oc^2$ and the potential energy $V$,
\begin{align}\label{2.8}
E = E'+m_oc^2+V.
\end{align}
Relativistic mass is little used by modern physicists. Notwithstanding the modern usage I have used $m$ for relativistic mass and $m_0$ for rest mass throughout the article.
In Eq.~\eqref{9.1} we replace $E$ by $i\hbar\partial/\partial t$ and $\boldsymbol{p}$ by an unconventional quantum mechanical operator $-i\hbar\boldsymbol{\nabla}_j$. Assuming spherically symmetric potential, we define operator $\boldsymbol{\nabla}_j^2$ in spherical polar coordinates as
\begin{align}\label{2.21}
\boldsymbol{\nabla}_j^2=\frac{1}{r^2}\left[\frac{\partial}{\partial r}\left(r^2\frac{\partial}{\partial r}\right)-\frac{1}{\hbar^2}(\boldsymbol{L}+\boldsymbol{S})^2\right], 
\end{align}
where $\boldsymbol{L}$ is the orbital angular momentum operator, and
$\boldsymbol{S}$ is the spin angular momentum operator. For electron, $\boldsymbol{S}=\frac{1}{2}\hbar\boldsymbol{\sigma}$ where $\boldsymbol{\sigma}$ are the Pauli spin matrices \cite{11}. Here the total angular momentum operator $\boldsymbol{J}$ is given by
\begin{align}\label{2.41}
 \begin{split}
\boldsymbol{J}&=(\boldsymbol{L}+\boldsymbol{S}).
 \end{split}
\end{align}
where spin corresponds to intrinsic angular momentum of the point particle.

If we put $\boldsymbol{S}=0$ in Eq.~\eqref{2.21}, we get the classical Laplacian $\boldsymbol{\nabla}^2$ of the Schr$\ddot{o}$dinger theory. Operator $\boldsymbol{\nabla}_j^2$ is associated with the total angular momentum of the particle and is capable of replacing Dirac matrices. Introduction of this operator in Eq.~\eqref{9.1} yields four spinors. The rectangular coordinates in terms of the spherical polar coordinate system we are using, have been defined according to the scheme $x=r\sin\theta\cos\phi$, $y=r\sin\theta\sin\phi$, and $z=r\cos\theta$. The explicit form of the momentum operators $L_x$, $L_y$, $L_z$, and $J_x$, $J_y$, $J_z$ remain same as the standard theory with same physical interpretation.

The Laplacian in spherical coordinates contains the squared radial momentum operator $p_r^2$ which is Hermitian. However, as pointed out by several authors \cite{9,74,75,79} the radial momentum operator $p_r$ of Dirac theory is not Hermitian. The proposed total momentum operator $\boldsymbol{\nabla}_j^2$ also contains the squared radial momentum operator $p_r^2$ which is Hermitian. The angular momentun operator $\boldsymbol{L}$, and the spin operator $\boldsymbol{S}$ contributes nothing to the the squared radial momentum operator $p_r^2$. Therefore in this respect the theory has an advantage over Dirac theory.

The wave equation~\eqref{9.1} can be written as
\begin{align}\label{9.2}
i\hbar\frac{\partial \psi}{\partial t} =\pm[-\hbar^2c^2\boldsymbol{\nabla}_j^2+(m_oc^2)^2]^{1/2}\psi+V\psi,
\end{align}
The wave equation~\eqref{9.2} can be considerably simplified if potential V does not depend on time. It is then possible to express its general solution as a sum of products of functions of $\boldsymbol{r}$ and $t$ separately. We consider a particular solution of Eq.~\eqref{9.2} that can be written as a product $\psi(\boldsymbol{r},t) = z(\boldsymbol{r}) f(t)$. A general solution can be written as a sum of such separated solutions. If we substitute the above product in Eq.~\eqref{9.2} and divide thru by the product, we get
\begin{align}\label{2.42}
i\hbar\frac{1}{f}\frac{df}{dt}=\pm\frac{1}{z}\left[-\hbar^2c^2\boldsymbol{\nabla}_j^2+(m_oc^2)^2\right]^{1/2}z+V.
\end{align}
If we define another function $u(\mathbf{r})$ such that
\begin{align}\label{2.43}
\frac{1}{z}\left[1-\left(\frac{\hbar}{m_0c}\right)^2\boldsymbol{\nabla}_j^2\right]^{1/2}z=\left[1-\left(\frac{\hbar}{m_0c}\right)^2\frac{1}{u}\boldsymbol{\nabla}_j^2u\right]^{1/2},
\end{align}
then we can write Eq.~\eqref{2.42} as
\begin{align}\label{2.17}
i\hbar\frac{1}{f}\frac{df}{dt}=\pm\left[-\frac{\hbar^2c^2}{u}\boldsymbol{\nabla}_j^2u+(m_oc^2)^2\right]^{1/2}+V.
\end{align}
Since the left side depends only on $t$ and right side only on $\boldsymbol{r}$, both side must be equal to a same separation constant which in this case is $(E+K)$ where  
\begin{align}\label{2.44}
K=(E-V)\left[-1+\left\{1+\frac{\kappa\alpha\hbar^2c^2}{2(E-V)^2}\right\}^{1/2}\right].
\end{align}
$K$ represents correction to the particle energy level due to the orbit curvature $\kappa$. Energy levels of two particles having identical parameters (including velocity), and one having a straight trajectory of a free particle and another having a curvature of the bound orbit, cannot be the same. This results in different relativistic masses and therefore different energy levels. The curvature of the orbit can alter the relativistic mass without changing the velocity. Hence we can define the total relativistic mass $m_t$ of the particle as 
\begin{align}\label{2.45}
(E-V)=m\:c^2+m_kc^2=(m+m_k)c^2=m_tc^2.
\end{align}
$\alpha$ is a function of the particle energy which we define as  
\begin{align}\label{2.23}
\alpha=\frac{2\{(m_oc^2)^2-E^2\}^\frac{1}{2}}{\hbar c}.
\end{align}
In case of circular orbits $\kappa=1/r$, and for a free particle $\kappa=0$ and thus $K=0$. Then equation for $f$ can be easily integrated to give,
\begin{align}\label{2.18}
f(t)=C\:\exp(-i(E+K)t/\hbar), 
\end{align}
where $C$ is an arbitrary constant and the equation for $u$ becomes,
\begin{align}\label{2.20}
\{(E-V)^2-(m_oc^2)^2\}u(\boldsymbol{r})=-\hbar^2c^2(\boldsymbol{\nabla}_j^2+\kappa\alpha/2)u(\boldsymbol{r}). 
\end{align}
If we expand both sides of Eq.~\eqref{2.43}, we find that for slow moving particles, $z\approx u$.

\subsubsection{Separation of angular and radial wave equations}

\hspace*{5 mm}Substitution of Eq.~\eqref{2.21} in  Eq.~\eqref{2.20} gives Eq.~\eqref{2.25}.
The radial and the angular parts are then separated by substituting
\begin{align}\label{2.24}
u(r,\theta,\phi)=R(r)Y(\theta,\phi)
\end{align}
in Eq.~\eqref{2.25} and dividing thru by RY. As is expected in the presence of spin, the radial and the angular wave functions do get split into two spinor components. These two components are sufficient to explain both positive and negative enenrgies. 

\begin{align}\label{2.25}
 \begin{split}
&\left[\frac{1}{\hbar^2c^2}\{(m_oc^2)^2-(E-V)^2\}-\frac{\alpha}{2r}\right]u(\boldsymbol{r})\\&=\left[\frac{1}{r^2}\frac{\partial}{\partial r}\left(r^2\frac{\partial}{\partial r}\right)-\left[i\left(\frac{1}{r^2\sin\theta}\frac{\partial}{\partial \theta}\left(\sin\theta\frac{\partial}{\partial \theta}\right)+\frac{1}{r^2\sin^2\theta}\frac{\partial^2}{\partial\phi^2}\right)^\frac{1}{2}+\frac{\boldsymbol{S}}{r\hbar}\right]^2\right]u(\boldsymbol{r}).
 \end{split}
\end{align}
\begin{align}\label{2.26}
 \begin{split}
&\frac{1}{R}\frac{d}{dr}\left(r^2\frac{dR}{dr}\right)-\frac{r^2}{\hbar^2c^2}\{(m_oc^2)^2-(E-V)^2\}+\frac{\alpha r}{2}\\ &=\frac{1}{Y}\frac{1}{\hbar^2}\left[i\hbar\left(\frac{1}{\sin\theta}\frac{\partial}{\partial \theta}\left(\sin\theta\frac{\partial Y}{\partial \theta}\right)+\frac{1}{\sin^2\theta}\frac{\partial^2Y}{\partial\phi^2}\right)^\frac{1}{2}+\boldsymbol{S}\sqrt{Y}\right]^2.
 \end{split}
\end{align}
Since the left side of Eq.~\eqref{2.26} depends only on $r$ and the right side depends only on $\theta$ and $\phi$, both sides must be equal to a constant that we call $\Gamma$. Thus Eq.~\eqref{2.26} gives us a radial equation 
\begin{align}\label{2.27}
\frac{1}{r^2}\frac{d}{dr}\left(r^2\frac{dR}{dr}\right)-\frac{1}{\hbar^2c^2}\{(m_oc^2)^2-(E-V)^2\}R
+\frac{\alpha}{2r}R-\frac{\Gamma}{r^2}R=0,
\end{align}
and an angular equation
\begin{align}\label{2.28}
\frac{1}{\hbar^2}\biggl[i\hbar\left(\frac{1}{\sin\theta}\frac{\partial}{\partial \theta}\left(\sin\theta\frac{\partial Y}{\partial \theta}\right)+\frac{1}{\sin^2\theta}\frac{\partial^2Y}{\partial\phi^2}\right)^\frac{1}{2}
+\boldsymbol{S}\sqrt{Y}\biggr]^2-\Gamma Y=0. 
\end{align}
The angular equation can be further separated by substituting 
\begin{align}\label{2.29}
Y(\theta,\phi)=\Theta(\theta)\Phi(\phi)
\end{align}
and dividing by $\Theta\Phi$.
\begin{align}\label{2.30}
\biggl[\frac{1}{\Theta}\frac{1}{\sin\theta}\frac{d}{d\theta}\left(\sin\theta\frac{d\Theta}{d\theta}\right)+\left(\sqrt{\Gamma}-\frac{\boldsymbol{S}}{\hbar}\right)^2\biggr]\sin^2\theta
=-\frac{1}{\Phi}\frac{d^2\Phi}{d\phi^2}=\nu.
\end{align}
From the theory of unitary groups and infinitesimal rotations \cite{11}, it follows that the spin angular momentum operator $\boldsymbol{S}^2$ is strictly a constant of the motion and can be replaced by the number $s(s+1)\hbar^2$ where $s$ is an integer or half an odd integer and we have already defined $\Gamma$ as a constant. Therefore the quantity $\{\sqrt{\Gamma}-(\boldsymbol{S}/\hbar)\}^2$ appearing in Eq.~\eqref{2.30} can only be a constant which we shall call $\lambda$. Hence we end up with two equations
\begin{align}\label{2.31}
\frac{d^2\Phi}{d\phi^2}+\nu\Phi=0,
\end{align}
\begin{align}\label{2.32}
\frac{1}{\sin\theta}\frac{d}{d\theta}\left(\sin\theta\frac{d\Theta}{d\theta}\right)+\left(\lambda-\frac{\nu}{\sin^2\theta}\right)\Theta=0.
\end{align}
Eqs.~\eqref{2.31} and ~\eqref{2.32} have exactly the same solutions as that given by Schr$\ddot{o}$dinger theory where $\nu$ is chosen to be equal to square of an integer $m$ which takes on positive or negative integer values or zero. Therefore,
\begin{align}\label{2.33}
\Phi_m(\phi)=(2\pi)^{-\frac{1}{2}}\:\exp(im\phi).
\end{align}
The condition that the solution of Eq.~\eqref{2.32} be finite at $\cos\theta=\pm1$ limits the values of $\lambda$ to $l(l+1)$ where $l$ is a positive integer or zero. The physically acceptable solutions of Eq.~\eqref{2.32} when $m=0$ are the Legendre polynomials and for $m\leq l$, the associated Legendre functions. 
Solution for $\lambda$ gives us the new relationship
\begin{align}\label{2.34}
\{\sqrt{\Gamma}-(\boldsymbol{S}/\hbar)\}^2=l(l+1)=\boldsymbol{L}^2/\hbar^2.
\end{align}
This can be written as
\begin{align}\label{2.35}
\{\sqrt{\Gamma}\hbar-\boldsymbol{S}\}^2=\boldsymbol{L}^2.
\end{align}
This condition can be satisfied if we substitute $\sqrt{\Gamma}\hbar=(\boldsymbol{L}+\boldsymbol{S})$ and this will also be consistent with Eq.~\eqref{2.28} which can be written as
\begin{align}\label{2.36}
(\boldsymbol{L}+\boldsymbol{S})^2Y=\Gamma\hbar^2Y.
\end{align}
Furthermore, the theory of unitary groups and infinitesimal rotations \cite{11} would require that we identify $\Gamma\hbar^2$ with eigenvalues $j(j+1)\hbar^2$ of the total angular momentum operator $\boldsymbol{J}^2$, where $j$ is zero or a positive integer or half an odd integer. This is acceptable for the radial Eq.~\eqref{2.27} and also for Eq.~\eqref{2.36} for integer $j=l$, but this will not permit seperation of variables in Eq.~\eqref{2.29} in case of hydrogen atom with spin 1/2 electron. Therefore Eqs.~\eqref{2.28} thru~\eqref{2.36} are valid only for integer values of $j$ and in a special case of $\boldsymbol{S}=0$ we get $\Gamma=\lambda=l(l+1)$ The angular part $Y_{lm}(\theta,\phi)$ of the complete wave function which is a solution of Eq.~\eqref{2.28} when $\Gamma=\lambda=l(l+1)$ is the spherical harmonic. If we put $\Gamma=\lambda=l(l+1)$ and $\kappa\alpha/2=0$ in the radial equation Eq.~\eqref{2.27}, we get the energy eigenvalues of the relativistic Schr$\ddot{o}$dinger theory. All these solutions of the Schr$\ddot{o}$dinger theory remain unaltered in this theory.

\subsubsection{Separation of angular wave equation for spin 1/2 electron}

In case of hydrogen with spin 1/2 electron, the separation of angular wave equation \cite{59} is carried out in the same manner as in Dirac theory. This can be done by redefining Eq.~\eqref{2.24} and Eq.~\eqref{2.36} as 
\begin{align}\label{2.36a}
\psi(r,\theta,\phi)=e^{-iEt/\hbar}R_{n,j}(r)\psi^{(2)}_{j,m_j}(\theta,\phi)=
\begin{bmatrix}
\psi_1\\
\psi_2
\end{bmatrix}
.
\end{align}
\begin{align}\label{2.36b}
(\boldsymbol{L}+\boldsymbol{S})^2\psi^{(2)}_{j,m_j}=\Gamma\hbar^2\psi^{(2)}_{j,m_j}.
\end{align}
Here common subscripts $l,1/2$ are omitted.
\begin{align}\label{2.36c}
\psi^{(2)}_{l+1/2,m+1/2}=\left(\frac{l+m+1}{2l+1}\right)^{1/2}\psi^{(1)}_{m,1/2} +\left(\frac{l-m}{2l+1}\right)^{1/2}\psi^{(1)}_{m+1,-1/2},
\end{align}
\begin{align}\label{2.36d}
\psi^{(2)}_{l-1/2,m+1/2}=\left(\frac{l-m}{2l+1}\right)^{1/2}\psi^{(1)}_{m,1/2} -\left(\frac{l+m+1}{2l+1}\right)^{1/2}\psi^{(1)}_{m+1,-1/2}.
\end{align}
Coefficients in Eqs.~\eqref{2.36c} and~\eqref{2.36d} are the Clebsch-Gordon coefficients \cite{60}. Here simultaneous eigenstates of $L^2$, $S^2$, $L_z$ and $S_z$ are given by
\begin{align}\label{2.36e}
\psi^{(1)}_{m,\pm1/2}=Y_{lm}(\theta,\phi)\chi_{\pm},
\end{align}
where $Y_{lm}(\theta,\phi)$ are the spherical harmonics from Eq.~\eqref{2.36} for $\boldsymbol{S}=0$, and $\chi_{\pm}$ are the unit spinors.

\subsubsection{Orbit curvature affects momentum}

\hspace*{5 mm} The radial wave equation Eq.~\eqref{2.27} may be rewritten in a form that resembles the classical Schr$\ddot{o}$dinger wave equation. If we put $R(r)=\chi(r)/r$ and $E^*=E'+V$ with reference to Eq.~\eqref{2.8}, then the equation for the modified radial wave function $\chi$ may be written as
\begin{align}\label{2.37}
 \begin{split}
&\frac{-\hbar^2}{2m_0}\frac{d^2\chi}{dr^2}+\biggl[V(\boldsymbol{r})-\frac{\hbar c}{r}\left(\frac{-E^*}{2m_0c^2}\right)^\frac{1}{2}-\frac{(E^*-V)^2}{2m_0c^2}\\
&+\frac{l(l+1)\hbar^2}{2m_0r^2}+\frac{s(s+1)\hbar^2}{2m_0r^2}+\frac{\hbar\boldsymbol{\sigma}\boldsymbol{.L}}{2m_0r^2}\biggr]\chi=E^*\chi.
 \end{split}
\end{align}
Thus the radial motion is similar to the classical motion of a particle in a potential defined by the quantity within the large brackets. The first term is equivalent to the classical kinetic energy $\boldsymbol{p}^2/2m_0$. The first, the second and the fifth terms are same as the classical Schr$\ddot{o}$dinger theory. The fourth term is approximately equal to $\boldsymbol{p}^4/8m_0^3c^2$ which has the form of the classical relativistic mass correction and is same as the Dirac theory. The seventh term includes spin-orbit coupling. The sixth term is due to the spin itself. The third term is associated with the curvature of the circular orbit $\kappa=1/r$ which originates from the separation constant $(E+K)$ given by Eq.~\eqref{2.44}. The magnitude of this term is dependent on the kinetic and potential energies in addition to the curvature. For very large $r$ this term becomes negligible. Therefore there is a certain amount of correction to the potential energy when a particle travels along an orbit having large curvature. There is a corresponding change in momentum. This curvature term contributes only to the relativistic mass of the particle and can be conviniently analyzed by substituting $-E^*=m_0c^2-E$. The effect of introducing the separation constant $(E+K)$ is equivalent to adding the term $\kappa\alpha/2$ to the operator $\boldsymbol{\nabla}_j^2$, provided $f(t)$ is defined by Eq.~\eqref{2.18}.

In Dirac's analysis, the third, fifth and sixth terms do not appear at all and the seventh spin orbit coupling term appear in the form of spin-orbit interaction energy associated with Larmor precession and Thomas precession \cite{2}. It also includes one more term as a correction to the potential energy and is declared difficult to demonstrate experimentally.

\subsubsection{Spin-orbit interaction energy}

\hspace*{5 mm} In our case, the last term in brackets in Eq.~\eqref{2.37} has nothing to do with the spin-orbit interaction energy but the last three terms together represent additional potential energy due to total angular momentum. Spin-orbit interaction energy term is hidden within the third term which is due to curvature effect and can be demonstrated as follows. It is to be noted that the parameter $1/r$ in this term does not have its origin in the classical Laplacian. $\gamma$ in the following expression is the fine structure constant. 
\begin{align}\label{2.38}
 \begin{split}
-\frac{\hbar c}{r}\left(\frac{-E^*}{2m_0c^2}\right)^\frac{1}{2}=\frac{V}{\sqrt{2}\gamma}\left(1-\frac{m_t}{m_0}-\frac{V}{m_0c^2}\right)^\frac{1}{2}.
 \end{split}
\end{align}
Using Eq.~\eqref{9.2} we can write
\begin{align}\label{2.38a}
(E-V)^2=(m_tc^2)^2=-\hbar^2c^2(\boldsymbol{\nabla}_j^2+\kappa\alpha/2)+(m_oc^2)^2,
\end{align}
\begin{align}\label{2.38b}
\frac{m_t}{m_0}=\pm\left[1-\left(\frac{\hbar}{m_0c}\right)^2(\boldsymbol{\nabla}_j^2+\kappa\alpha/2)\right]^{1/2}.
\end{align}
Eq.~\eqref{2.38b} shows that the particles can have positive or negative relativistic mass but the rest mass is always positive. This is the difference between particle and anti-particle. Since the second term in the bracket is small we can have
\begin{align}\label{2.38c}
\frac{m_t}{m_0}=\pm\left[1-\frac{1}{2}\left(\frac{\hbar}{m_0c}\right)^2(\boldsymbol{\nabla}_j^2+\kappa\alpha/2)\right].
\end{align}
Using Eq.~\eqref{2.21} we can write
\begin{align}\label{2.39}
\boldsymbol{\nabla}_j^2=\boldsymbol{\nabla}^2-\frac{1}{\hbar^2r^2}(\boldsymbol{S}^2+2\boldsymbol{L}\cdot\boldsymbol{S}).
\end{align}
Substitution of Eq.~\eqref{2.39} in Eq.~\eqref{2.38c} gives
\begin{align}\label{2.39a}
\frac{m_t}{m_0}=\pm\left[1+\frac{\boldsymbol{p}^2r^2+\boldsymbol{S}^2+2\boldsymbol{L}\cdot\boldsymbol{S}-(\alpha r\hbar^2/2)}{2m_0^2r^2c^2}\right].
\end{align}
For a free particle we have $\kappa\alpha/2=0$ and in the absence of spin $\boldsymbol{S}=0$. This reduces Eq.~\eqref{2.38b} to Lorentz transformation equation,
\begin{align}\label{2.39b}
\frac{m}{m_0}=\pm\left[1-\frac{v^2}{c^2}\right]^{-1/2}.
\end{align}
Next we substitute Eq.~\eqref{2.39a} with positive sign for positive energy in Eq.~\eqref{2.38} to obtain 
\begin{align}\label{2.40}
 \begin{split}
-\frac{\hbar c}{r}&\left(\frac{-E^*}{2m_0c^2}\right)^\frac{1}{2}\approx\frac{V^\frac{1}{2}}{\gamma}\biggl(\frac{V}{4}\biggl\{-\frac{\boldsymbol{p}^2 r^2+\boldsymbol{S}^2-(\alpha r\hbar^2/2)}{m_0^2c^2r^2}\biggr\}\\
&-\frac{1}{2m_0^2c^2}\frac{1}{r}\frac{dV}{dr}\boldsymbol{L}\cdot\boldsymbol{S}-\frac{V^2}{2m_0c^2}\biggr)^\frac{1}{2}.
 \end{split}
\end{align}
The middle term on the right has the form of the spin-orbit interaction energy associated with Larmor precession and Thomas precession \cite{11,2}.

Now we have sufficient information to conclude that the term $\kappa\alpha/2$ appearing in Eq.~\eqref{2.44} simulates the bound state of the electron and provides coupling between the momentum, orbit curvature, spin-orbit interaction and the central coulomb potential. Therefore in order to deal with the free particle, we have to put curvature $\kappa=0$, but this does not eliminate the spin and so it makes its presence felt in the presence of an external electromagnetic field as in case of the Pauli equation \cite{61}.

\section{The Hydrogen atom}
\subsection{Energy levels}

The second term in radial equation ~\eqref{2.27} is,
\begin{align}\label{3.1}
	\begin{split}
-\frac{1}{\hbar^2c^2}\{(m_oc^2)^2-(E-V)^2\}R=
-\frac{(m_oc^2+E-V)}{\hbar c}\frac{(m_oc^2-E+V)}{\hbar c}R.
	\end{split}
\end{align}
We will substitute 
\begin{align}\label{3.2}
\alpha_1=\frac{2(m_oc^2+E)}{\hbar c},\ and\ \alpha_2=\frac{2(m_oc^2-E)}{\hbar c}.
\end{align}
Hence with respect to Eq.~\eqref{2.23}, $\alpha^2=\alpha_1\alpha_2$. The attractive coulomb interaction between an atomic nucleus of charge $+Ze$ and an electron of charge $-e$ is represented by the potential energy $V(r)=-Ze^2k'/r$ where $k'$ is Coulomb's constant. In case of hydrogen atom, $Z=1$. We introduce these hydrogen parameters into Eq.~\eqref{3.1} with the fine structure constant $\gamma$ defined as $\gamma=(e^2k'Z)/(\hbar c)$.
\begin{align}\label{3.3}
	\begin{split}
-\frac{1}{\hbar^2c^2}\{(m_oc^2)^2-(E-V)^2\}R=
-\alpha^2\left(\frac{1}{4}+\frac{\gamma}{r}\left(\frac{\alpha_2-\alpha_1}{2\alpha^2}\right)-\frac{\gamma^2}{r^2\alpha^2}\right)R
	\end{split}
\end{align}
We will rewrite the radial equation ~\eqref{2.27} in dimensionless form by introducing a unitless independent variable $\rho=\alpha r$. Substitution of Eqs.~\eqref{3.3} and ~\eqref{2.36} gives,
\begin{align}\label{3.4}
 \begin{split}
\frac{1}{\rho^2}\frac{d}{d\rho}\left(\rho^2\frac{dR}{d\rho}\right)-\left(\frac{1}{4}+\frac{\gamma}{\rho}\left(\frac{\alpha_2-\alpha_1}{2\alpha}\right)-\frac{\gamma^2}{\rho^2}\right)R
+\frac{1}{2\rho}R-\frac{j(j+1)}{\rho^2}R=0,
 \end{split}
\end{align}
As far as the leading terms are concerned, for sufficiently large $\rho$ it is apparent that $R(\rho)=\rho^n e^{\pm\frac{1}{2}\rho}$ satisfies Eq.~\eqref{3.4} when $n$ has any finite value. This suggests that we look for an exact solution of Eq.~\eqref{3.4} of the form
\begin{align}\label{3.5}
R(\rho)=F(\rho)e^{-\frac{1}{2}\rho}
\end{align}
where $F(\rho)$ is a polynomial of finite order in $\rho$. Substitution of Eq.~\eqref{3.5} into Eq.~\eqref{3.4} gives equation for $F(\rho)$ as 
\begin{align}\label{3.6}
F"+\left(\frac{2}{\rho}-1\right)F'+\biggl[\frac{\lambda-(1/2)}{\rho}-\frac{j(j+1)-\gamma^2}{\rho^2}\biggr]F=0,
\end{align}
where we have substituted
\begin{align}\label{3.7}
\lambda=-\gamma\left(\frac{\alpha_2-\alpha_1}{2\alpha}\right).
\end{align}
Now we find a solution for $F$ in the form
\begin{align}
F(\rho)=\rho^s(a_0+a_1\rho+a_2\rho^2+\cdots)=\rho^sL(\rho),\quad
\ a_0\neq 0,\ s\geq 0.\label{3.8}
\end{align}
Substitution of Eq.~\eqref{3.8} into Eq.~\eqref{3.6} gives us the equation for $L$.
\begin{align}\label{3.10}
	\begin{split}
\rho^2L"+\rho\{2(s+1)-\rho\}L'+[\rho\{\lambda-s-(1/2)\}
+s^2+s+\gamma^2-j(j+1)]L=0.
	\end{split}
\end{align}
If we set $\rho=0$ in Eq.~\eqref{3.10}, it follows from Eq.~\eqref{3.8} that
\begin{align}\label{3.11}
s^2+s+\gamma^2-j(j+1)=0.
\end{align}
This quadratic equation in $s$ has two solutions,
\begin{align}\label{3.12}
s=-\frac{1}{2}\pm\left\{\biggl(j+\frac{1}{2}\biggr)^2-\gamma^2\right\}^\frac{1}{2}.
\end{align}
The boundary condition that $R(\rho)$ be finite at $\rho=0$ requires that we choose upper sign for $s$. It is to be noted here that both, Schr$\ddot{o}$dinger's relativistic theory as well as Dirac's theory allow value of $s$ which is very slightly less than the permissible value. This problem does not exist in this theory. The smallest value in this theory is $\approx(1-\gamma^2)/2$, well within the range $s\geq 0$. With this, Eq.~\eqref{3.10} reduces to
\begin{align}\label{3.13}
\rho L"+\{2(s+1)-\rho\}L'+\{\lambda-s-(1/2)\}L=0.
\end{align}
Equation ~\eqref{3.13} can be solved by substituting Eq.~\eqref{3.8}. The recursion relation between the coefficients of successive terms of the series is observed to be
\begin{align}\label{3.14}
a_{\nu+1}=\frac{(\nu+s+(1/2)-\lambda)}{\nu(\nu+1)+2(\nu+1)(s+1)}\:a_\nu.
\end{align}
If the series does not terminate, its dominant asymptotic behavior when $\rho\rightarrow\infty$ can be inferred from the coefficients of its high terms:
\begin{align}\label{3.15}
\frac{a_{\nu+1}}{a_\nu}\rightarrow\frac{1}{\nu}.
\end{align}
This means that the series has the asymptotic form $e^{\rho}$ and regular solution is obtained only if it terminates. Suppose that this occurs at $\nu=n'$, so that $a_{n'+1}=0$. Then Eq.~\eqref{3.14} give the relation
\begin{align}\label{3.16}
n'+s+(1/2)-\lambda=0, \hspace*{5 mm}  n'=0,1,2,\ldots
\end{align}
Substituting for $s$ and $\lambda$ from Eqs.~\eqref{3.12} and ~\eqref{3.7} respectively, we get the energy levels of hydrogen atom
\begin{align}\label{3.17}
E=m_0c^2\left(1+\frac{\gamma^2}{\left[\left\{\left(j+\frac{1}{2}\right)^2-\gamma^2\right\}^\frac{1}{2}+n'\right]^2}\right)^{-\frac{1}{2}},
\end{align}
where the radial quantum number $n'$ is related to the total quantum number $n$ by the expression
\begin{align}\label{3.18}
n=n'+(j+(1/2)).
\end{align}
These are exactly the same Dirac energy levels having the same total spread in energy of the fine-structure levels for a given $n$.

\subsection{Radial wave function}

It is clear that Eq.~\eqref{3.13} has the form of associated Laguerre differential equation \cite{11},
\begin{align}\label{3.18.1}
\rho L^{p}_{q}\:"+(p+1-\rho)L^{p}_{q}\:'+(q-p)L^{p}_{q}=0.
\end{align}
The associated Laguerre polynomials $L^{p}_{q}$ can be constructed according to formula
\begin{align}\label{3.18.2}
L^{p}_{q}\:(\rho)=\sum^{q-p}_{k=0}\:(-1)^{k+p}\frac{[q!]^2\rho^k}{(q-p-k)!(p+k)!k!}.
\end{align}
In case of Eq.~\eqref{3.13}, we have $p=2s+1$, $(q-p)=\lambda-s-(1/2)$ and $q=\lambda+s+(1/2)$. Comparision with Eq.~\eqref{3.16} shows that $(q-p)$ are integers but $p$ and $q$ are not integers. Therefore it is possible to solve Eq.~\eqref{3.13} using Laguerre polynomials only if we introduce the approximations $p\approx2j+1$ and $q\approx\lambda-s+2j+(1/2)$ by ignoring $\gamma^2$ term in $s$. Hence we get,
\begin{align}\label{3.18.3}
 \begin{split}
L^{2j+1}_{\lambda-s+2j+(1/2)}\:(\rho)=
\sum^{\lambda-s-(1/2)}_{k=0}\:(-1)^{k+2j+1}\frac{[(\lambda-s+2j+(1/2))!]^2\rho^k}{(\lambda-s-(1/2)-k)!(2j+1+k)!k!}.
 \end{split}
\end{align}
This will yield two associated Laguerre polynomials $L^{p}_{q}$ corresponding to two values of $\lambda$ associated with two spinors which are applicable to positive energy as well as the negative energy solutions. The resulting approximate radial wave function is of the form $e^{-\frac{1}{2}\rho}\rho^s\:L^{2j+1}_{\lambda-s+2j+(1/2)}$ and the normalization constant may be found by using the generating function to evaluate the integral
\begin{align}\label{3.18.4}
\int^{\infty}_{0}e^{-\rho}\rho^{2s}\:[L^{2j+1}_{\lambda-s+2j+(1/2)}(\rho)]^2\rho^2d\rho.
\end{align}

\section{The quarkonium extension}

The theory of the preceding sections can be extended to quarkonium physics \cite{62} as follows. We use the relativistic hamiltonian of the form
\begin{align}\label{4.1}
H=H_0+V_0(r)
\end{align}
where $H_0$ is the relativistic kinetic energy \cite{65,66,67} given by
\begin{align}\label{4.2}
H_0=\sqrt{p^2+m_1^2}+\sqrt{p^2+m_2^2}.
\end{align}
Here $m_1$, $m_2$ are masses of quark antiquark, $p$ the momentum and the units are so chosen that $\hbar=c=1$. $V_0(r)$ is chosen to be the one-gluon-exchange Coulomb plus linear potential also known as the Cornell potential \cite{63,64} given by
\begin{align}\label{4.3}
V_0(r)=-\frac{4\alpha_s}{3r}+\sigma r+D.
\end{align}
Where $\alpha_s$ is the QCD effective coupling constant, $\sigma$ the string tension and $D$ a constant.
Hence for charmonium $(c\bar{c})$ and bottomonium $(b\bar{b})$ we get the wave equation of the form
\begin{align}\label{4.4}
E\psi=\left\{2\sqrt{p^2+m_q^2}+V_0\right\}\psi.
\end{align}
\begin{align}\label{4.4a}
i\frac{\partial \psi}{\partial t} =2[-\boldsymbol{\nabla}_j^2+m_q^2]^{1/2}\psi+V_0\psi,
\end{align}
If we substitute $\psi(\boldsymbol{r},t) = z(\boldsymbol{r}) f(t)$ and divide thru by $zf$, we get
\begin{align}\label{4.4b}
i\frac{1}{f}\frac{df}{dt}=\frac{2}{z}\left[-\boldsymbol{\nabla}_j^2+m_q^2\right]^{1/2}z+V_0.
\end{align}
If we define another function $u(\boldsymbol{r})$ such that
\begin{align}\label{4.4c}
\frac{1}{z}\left[1-\frac{1}{m_q^2}\boldsymbol{\nabla}_j^2\right]^{1/2}z=\left[1-\frac{1}{m_q^2}\frac{1}{u}\boldsymbol{\nabla}_j^2u\right]^{1/2},
\end{align}
then we can write Eq.~\eqref{4.4b} as
\begin{align}\label{4.4d}
i\frac{1}{f}\frac{df}{dt}=2\left[-\frac{1}{u}\boldsymbol{\nabla}_j^2u+m_q^2\right]^{1/2}+V_0.
\end{align}
Since the left side depends only on $t$ and right side only on $\boldsymbol{r}$, both side must be equal to a same separation constant which in this case is $(E+K_q)$ where  
\begin{align}\label{4.4e}
K_q=(E-V_0)\left[-1+\left\{1+\frac{2\kappa\alpha}{(E-V_0)^2}\right\}^{1/2}\right].
\end{align}
We define $\alpha$ for $c\bar{c}$ and $b\bar{b}$ with the relation
\begin{align}\label{4.6}
\alpha=\frac{1}{2}\{E^2-4m_q^2\}^\frac{1}{2},
\end{align}
Equation for $f$ can be easily integrated to give,
\begin{align}\label{4.5}
f(t)=C\:\exp(-i(E+K_q)t), 
\end{align}
where $C$ is an arbitrary constant and the equation for $u$ becomes,
\begin{align}\label{4.7}
\boldsymbol{(\nabla}_j^2+\kappa\alpha/2)u(\boldsymbol{r})=\left\{m_q^2-\frac{1}{4}(E-V_0)^2\right\}u(\boldsymbol{r}).
\end{align}
Separation of radial and angular wave equations is accomplished by steps identical to Eqs.~\eqref{2.24},~\eqref{2.25} and~\eqref{2.26}. This gives us a radial equation of the form
\begin{align}\label{4.8}
 \begin{split}
\frac{1}{r^2}\frac{d}{dr}\left(r^2\frac{dR}{dr}\right)-\left\{m_q^2-\frac{1}{4}(E-V_0)^2\right\}R
+\frac{\alpha}{2r}R
-\frac{\Gamma}{r^2}R=0.
 \end{split}
\end{align}
The angular wave equation and its further separation is exactly same as Eqs.~\eqref{2.28} to ~\eqref{2.36e}. In radial equation we substitute
\begin{align}\label{4.9}
\alpha_1=\frac{1}{2}(E+2m_q),\ and\ \alpha_2=\frac{1}{2}(E-2m_q).
\end{align}
\begin{align}\label{4.10}
 \begin{split}
\frac{1}{\rho^2}\frac{d}{d\rho}\left(\rho^2\frac{dR}{d\rho}\right)+\left(1-\frac{V_0 E}{2\alpha^2}+\frac{V_0^2}{4\alpha^2}\right)R
+\frac{1}{2\rho}R-\frac{\Gamma}{\rho^2}R=0,
 \end{split}
\end{align}
Substitution of Eqs.~\eqref{4.3} and~\eqref{3.5} gives
\begin{align}\label{4.11}
 \begin{split}
F"&+\left(\frac{2}{\rho}-1\right)F'+\biggl[\frac{\sigma^2}{4\alpha^4}\rho^2+\frac{\sigma(D-E)}{2\alpha^3}\rho
+\left(\frac{4\alpha_s^2}{9}-\Gamma\right)\frac{1}{\rho^2}\\
&+\left(\frac{2\alpha_s(E-D)}{3\alpha}-\frac{1}{2}\right)\frac{1}{\rho}
+\left\{\frac{5}{4}+\frac{1}{2\alpha^2}\left(\frac{D^2}{2}-ED-\frac{4\alpha_s\sigma}{3}\right)\right\}\biggr]F=0.
 \end{split}
\end{align}
Substitution of Eq.~\eqref{3.8} and dividing thru by $\rho^{(s-2)}$ yields
\begin{align}\label{4.12}
	\begin{split}
\rho^2L"+\rho\{2(s+1)-\rho\}L'+[s(s+1)+(H-s)\rho
+G\rho^2+B\rho^3+A\rho^4+C]L=0.
	\end{split}
\end{align}
\begin{align}\label{4.13}
\text{where}\quad A=\frac{\sigma^2}{4\alpha^4},\quad B=\frac{\sigma(D-E)}{2\alpha^3}, \quad C=\left(\frac{4\alpha_s^2}{9}-\Gamma\right),
\end{align}
\begin{align}\label{4.14}
H=\left(\frac{2\alpha_s(E-D)}{3\alpha}-\frac{1}{2}\right), \quad G=\left\{\frac{5}{4}+\frac{1}{2\alpha^2}\left(\frac{D^2}{2}-ED-\frac{4\alpha_s\sigma}{3}\right)\right\}.
\end{align}
If we put $\rho=0$ in Eq.~\eqref{4.12} then we must have
\begin{align}\label{4.16}
s(s+1)+C=0,
\end{align}
\begin{align}\label{4.17}
s=-\frac{1}{2}+\left\{\biggl(j+\frac{1}{2}\biggr)^2-\frac{4\alpha_s^2}{9}\right\}^\frac{1}{2}.
\end{align}
This reduces Eq.~\eqref{4.12} to
\begin{align}\label{4.18}
	\begin{split}
\rho L"+\{2(s+1)-\rho\}L'+[A\rho^3+B\rho^2+G\rho
+(H-s)]L=0.
 \end{split}
\end{align}
Equation ~\eqref{4.18} can be solved by substituting Eq.~\eqref{3.8}. The recursion relation between the coefficients of successive terms of the series is observed to be
\begin{align}\label{4.19}
a_{\nu+1}=\frac{(\nu-A\rho^3-B\rho^2-G\rho-(H-s))}{\nu(\nu+1)+2(\nu+1)(s+1)}\:a_\nu.
\end{align}
Suppose that the series terminates at $\nu=n'$, so that $a_{n'+1}=0$. Then Eq.~\eqref{4.19} gives a cubic equation in $\rho$,
\begin{align}\label{4.20}
A\rho^3+B\rho^2+G\rho+(H-s-n')=0. \hspace*{5 mm}  n'=0,1,2,\ldots
\end{align}
Substitution of Eqs.~\eqref{4.13},~\eqref{4.14},~\eqref{4.17}, and $\rho=\alpha r$ gives
\begin{align}\label{4.21}
	\begin{split}
&r^3+\frac{2(D-E)}{\sigma}r^2+\frac{1}{\sigma^2}\left(5\alpha^2+D^2-2ED
-\frac{8\alpha_s \sigma}{3} \right) r +\\
&\frac{4}{\sigma^2}\left[\frac{2}{3}\alpha_s(E-D)
-\alpha \left[\left\{\left(j+\frac{1}{2}\right)^2
-\frac{4\alpha_s^2}{9}\right\}^{1/2} +n'\right]\right]=0	
  \end{split}
\end{align}
Substitution of following gives Eq.~\eqref{4.23}.
\begin{align}\label{4.22}
r=\left(r_1-\frac{2(D-E)}{3\sigma}\right),
\end{align}
\begin{align}\label{4.23}
r_1^3+ar_1+b=0, \quad \text{where}
\end{align}
\begin{align}\label{4.24}
a=\left[\frac{1}{\sigma^2}\left(5\alpha^2+D^2-2ED
-\frac{8\alpha_s \sigma}{3}\right)-\frac{4(D-E)^2}{3\sigma^2}\right],
\end{align}
\begin{align}\label{4.25}
	\begin{split}
&b=\frac{16}{27}\frac{(D-E)^3}{\sigma^3}-\frac{2(D-E)}{3\sigma^3}\left(5\alpha^2+D^2-2ED
-\frac{8\alpha_s \sigma}{3}\right)\\
&+\frac{4}{\sigma^2}\left[\frac{2}{3}\alpha_s(E-D)
-\alpha \left[\left\{\left(j+\frac{1}{2}\right)^2
-\frac{4\alpha_s^2}{9}\right\}^{1/2} +n'\right]\right]
  \end{split}
\end{align}
Cubic Eq.~\eqref{4.23} has the solution
\begin{align}\label{4.26}
r_1=\left(-\frac{b}{2}+d\right)^{1/3}+\left(-\frac{b}{2}-d\right)^{1/3}, \quad \text{where} \quad d=\sqrt{\frac{a^3}{27}+\frac{b^2}{4}}.
\end{align}
Eq.~\eqref{4.18} has the form of associated Laguerre differential equation~\eqref{3.18.1}. Therefore it follows that
\begin{align}\label{4.28}
p=2s+1, \quad (q-p)=A\rho^3+B\rho^2+G\rho+(H-s), 
\end{align}
\begin{align}\label{4.29}
\text{and} \quad q=A\rho^3+B\rho^2+G\rho+(H+s+1). 
\end{align}
From Eq.~\eqref{4.20} we get $(q-p)=n'$ = integer, but $p$ and $q$ are not integers. 
Therefore it is possible to solve Eq.~\eqref{4.18} using Laguerre polynomials only if we introduce the approximations $p\approx2j+1$ and $q=A\rho^3+B\rho^2+G\rho+(H-s+2j+1)$ by ignoring term containing $\alpha_s^2$ in $s$. Separation of the angular wave equation will be as for the spin 1/2 particle.

\section{Heavy quarkonium spectra}

Here we highlight the fundamental differences in the application of this theory to the hydrogen problem and the quarkonium problem. In case of hydrogen, the radial distance parameter $r$ does not appear in Eq.~\eqref{3.16} and therefore it has only a statistical interpretation and we can only compute its expectation value. 
In case of quarkonium however, $r$ appears in Eq.~\eqref{4.20} and it becomes possible to define $r$ even before defining the wave function. As a matter of fact the established value of $r$ defines the wave function. Therefore $r$ does not have a statistical interpretation. This does not mean violation of uncertainty principle because we are not predicting both energy level and the radial distance simultaneously but we are using the measured energy levels of two spin states as the input parameters to theoretically determine the radial distance. The results for the charmonium spectra are shown in Table~\ref{tab:Table1} and for the bottomonium spectra in Table~\ref{tab:Table2}. Here we can see how it is possible to predict the exact value of $r$. The string tension $\sigma$ is sensitive to variation in the radial distance $r$ and therefore it is possible to simultaneously predict both $r$ and $\sigma$ using itterative method which is based on the radial distance balancing procedure. In this model the experimentally observed quarkonium energy levels are used as input parameters and what is predicted is the radial distance $r$ and the string tension $\sigma$. So the conformance to the observed energy levels is $100\%$. The main criteria behind the itterative method is the fact that, between the two spin dependent states $(l+s)$ and $(l-s)$, there is only one common radial distance permissible and this radial distance is associated with a specific value of the string tension $\sigma$. Unlike other non-relativistic and semi-relativistic models, $\sigma$ in this relativistic model is a variable output parameter. We classify the quarkonium bound states with the label $n'l_j$ where $n'$ is the radial quantum number, letters S, P, D, F are used for the orbital angular momentum eigenvalue $l=0,1,2,3$ respectively, and total angular momentum eigenvalue $j$ can take on two values $l+s$ and $l-s$ where for quarks $s=1/2$. Here we clarify that our model developed in Sec.4 is valid only for quark systems having same spin states for both the quarks, i.e., either both quarks having $l+s$ state or both having $l-s$ state. The model becomes very complicated if one of the quark has $l+s$ state and another $l-s$ state.

The classification presented in Tables~\ref{tab:Table1} and~\ref{tab:Table2} is only one out of several configurations which are possible in this theory. All $n'P_{1/2}$ states can also be defined as $n'S_{1/2}$ states, however $S$ states do not occur in spin dependent pairs, so the radial distance for this state cannot be pinned down like other states. Besides, as can be seen from the tables, entire quarkonium spectra can be explained with $P$, $D$ or $F$ states. So it is likely that $S$ states may not exist at all. Secondly one can also change the radial quantum number $n'$ and get a different radial distance and a wave function for the same energy level. Similarly $P$ state can be changed to $D$ state or $F$ state. Such manipulations are possible only because $r$ and $\sigma$ are not measurable parameters. Severe restrictions on this model can be imposed only if the experimental physics can come up with a way of measuring the radial distance or the string tension. Due to the latitude and the flexibility in this theory, any newly discovered energy level can be easily accomodated in the model. This theory has presented several predictions for $r$ and $\sigma$, so what is required from the experimental physics is to come up with a way to measure these parameters.

One major difference between this model and other non-relativistic and semi-relativistic models is the size of the $q\bar{q}$ system. Generally this size range between 0.2 fm and 1 fm. In our model it can go as high as 10 fm. In this connection it is interesting to note that the $rms$ radii given by \cite{84,85} has some similarity to this model. Similarly, most models use a constant value of $\sigma$. Typically $\sigma=0.18$ GeV$^2$. In our case $\sigma$ range between 0.4 and 19. For these reasons it is not very meaningful to make any comparision with the Sommer scale \cite{68,69}. One of the lattice QCD study does point out that $\sigma$ may not be a constant, but some function of $r$ \cite{87}. Perhaps this is the only model which directly introduces the quark spin $s=1/2$ in the dynamics of motion. Rest of the models following the guidelines of perturbative and lattice QCD use the concept of total spin $S=0$ and $1$, which gets introduced as a perturbation to the spin independent kinetic energy of the system.
\begin{table}
	 \caption{Charmonium $(c\bar{c})$ results with $\alpha_s=0.\:3$, $m_q=1.25$ GeV, and $D=0$. $r$(fm)$\:= r/5.06763628483$. \label{tab:Table1}}
\hskip4pc\vbox{\columnwidth=26pc 
		\begin{tabular} {lllll}
		$n'l_j$ \hspace*{5 mm} &$E$(GeV) \hspace*{5 mm} &$Expt.$(MeV) \hspace*{2 mm}  &$r$(fm) \hspace*{10 mm} &$\sigma$(GeV$^2$)\\  \hline 
		$1P_{1/2}$ &$3.096916$ &$\pm0.011$ &$1.1050359$ &$1.028435$\\
		$1P_{3/2}$ &$2.9804$		&$\pm1.2$ &$\hspace{5mm}"$ &$\hspace{5mm}"$\\
		$2P_{1/2}$ &$3.5562$ &$\pm0.09$ &$1.3354474$ &$0.921835$\\
		$2P_{3/2}$ &$3.41476$  &$\pm0.35$ &$\hspace{5mm}"$ &$\hspace{5mm}"$\\
		$3P_{1/2}$ &$3.686093$ &$\pm0.034$ &$0.9149081$ &$1.46332$\\
		$3P_{3/2}$ &$3.52593$ &$\pm0.27$ &$\hspace{5mm}"$ &$\hspace{5mm}"$\\
		$4P_{1/2}$ &$3.6380$ &$\pm4.0$ &$1.1376934$ &$1.16552$\\
		$4P_{3/2}$ &$3.51066$ &$\pm0.07$ &$\hspace{5mm}"$ &$\hspace{5mm}"$\\
		$5P_{1/2}$ &$3.929$ &$\pm5.4$ &$0.6963006$ &$2.215$\\
		$5P_{3/2}$ &$3.7711$ &$\pm2.4$ &$\hspace{5mm}"$ &$\hspace{5mm}"$\\
		$6P_{1/2}$ &$4.153$ &$\pm3.0$ &$2.0569224$ &$0.67248$\\
		$6P_{3/2}$ &$4.039$ &$\pm1.0$ &$\hspace{5mm}"$ &$\hspace{5mm}"$\\
		$7P_{1/2}$ &$4.421$ &$\pm4.0$ &$1.5351580$ &$0.9896$\\
		$7P_{3/2}$ &$4.290$ &$-$ &$\hspace{5mm}"$ &$\hspace{5mm}"$\\
				\end{tabular}
				}
\end{table}

\begin{table}
	 \caption{Bottomonium $(b\bar{b})$ results with $\alpha_s=0.\:2$, $m_q=4.63$ GeV, and $D=0$. $r$(fm)$\:= r/5.06763628483$.\label{tab:Table2}}
\hskip4pc\vbox{\columnwidth=26pc
		\begin{tabular} {lllll}
		$n'l_j$\hspace*{5 mm} &$E$(GeV)\hspace*{5 mm} &$Expt.$(MeV)\hspace*{2 mm} &$r$(fm) \hspace*{10 mm}&$\sigma$(GeV$^2$)\\  \hline 
		$1P_{1/2}$ &$9.4603$ &$\pm0.26$ &$0.9311968$ &$3.987$\\
		$1P_{3/2}$ &$9.420$		&$-$ &$\hspace{5mm}"$ &$\hspace{5mm}"$\\
		$2P_{1/2}$ &$10.0236$ &$\pm0.31$ &$0.6721207$ &$5.73269$\\
		$2P_{3/2}$ &$9.91221$  &$\pm0.40$ &$\hspace{5mm}"$ &$\hspace{5mm}"$\\
		$2D_{3/2}$ &$9.91221$ &$\pm0.40$ &$4.9635941$ &$0.7583$\\
		$2D_{5/2}$ &$9.89278$ &$\pm0.40$ &$\hspace{5mm}"$ &$\hspace{5mm}"$\\
		$2F_{5/2}$ &$9.89278$ &$\pm0.40$ &$2.5353616$ &$1.49095$\\
		$2F_{7/2}$ &$9.85944$ &$\pm0.52$ &$\hspace{5mm}"$ &$\hspace{5mm}"$\\
		$3P_{1/2}$ &$10.26865$ &$\pm0.55$ &$9.9546316$ &$0.382645$\\
		$3P_{3/2}$ &$10.25546$ &$\pm0.55$ &$\hspace{5mm}"$ &$\hspace{5mm}"$\\
		$3D_{3/2}$ &$10.25546$ &$\pm0.55$ &$5.3828776$ &$0.709295$\\
		$3D_{5/2}$ &$10.2325$ &$\pm0.6$ &$\hspace{5mm}"$ &$\hspace{5mm}"$\\
		$3F_{5/2}$ &$10.2325$ &$\pm0.6$ &$1.2616777$ &$3.0839$\\
		$3F_{7/2}$ &$10.1611$ &$\pm1.7$ &$\hspace{5mm}"$ &$\hspace{5mm}"$\\
		$4P_{1/2}$ &$10.5794$ &$\pm1.2$ &$0.2311844$ &$18.6851$\\
		$4P_{3/2}$ &$10.3552$ &$\pm0.5$ &$\hspace{5mm}"$ &$\hspace{5mm}"$\\
				\end{tabular}
				}
\end{table}

\section{Conclusion}

Spin is successfully introduced in the new relativistic wave equation. The resulting energy levels of Hydrogen atom in this theory are exactly same as that of Dirac's theory. The separation of the new wave equation in spherical polar coordinates is as simple as that of the Schr$\ddot{o}$dinger theory and the solution of the angular part of the wave equation for spin 1/2 electron is same as in Dirac's theory. The curvature of the orbit is intorduced by proposing a novel form of separation constant for the bound state radial wave equation, which reduces to standard value $E$ for a free particle. The radial wave function of the Schr$\ddot{o}$dinger theory gets split into two spinor components. These two components are sufficient to explain both positive and negative energies. The new theory accounts for the energy due to spin-orbit interaction as well as for the additional potential energy due to spin and spin-orbit coupling. Spin angular momentum operator $\boldsymbol{S}$ is as neatly integrated into the equation of motion as the orbital angular momentum operator $\boldsymbol{L}$. Consequently the Dirac matrices and the k operator of Dirac's theory are dispensed with. The theory also points out that the curvature of the orbit draws on certain amount of kinetic and potential energies affecting the momentum of electron and the spin-orbit interaction energy constitutes a part of this energy. It is shown \cite{61} that the correction introduced to the radial equation due to the modification of classical Laplacian and the introduction of curvature is not the same as that of the Darwin and Pauli theroy however, the resulting energy levels of Hydrogen atom are same as that of Darwin, Pauli and Dirac theories. This is because the new theory is not a perturbation theory. Dirac's theory cannot detect the curvature effect because spin in that theory is not a part of the dynamics of motion but it gets introduced as a perturbation just like Darwin and Pauli theory. Also the selection of the radial momentum operator is somewhat arbitrary in that theory \cite{14,15,16,17,18} and not Hermitian as pointed out by several authors. The probability density and the current density for the hydrogen atom have been discussed elsewhere \cite{61}. The probability density and the current density for a free particle is comparable to that of the Pauli equation for the motion of electron in an electromagnetic field with spin included. In the presence of the external electromagnetic field, the theory reduces to the Pauli equation in the non relativistic limit \cite{61}. Comparision of the new wave equation with the Klein-Gordon equation shows why the later cannot account for spin values other than zero \cite{61}. 

The phenomenological non-perturbative, non-linear and relativistic quarkonium model presented here is unique in the sense that the radial distance in the quark model does not have a statistical interpretation like the hydrogen atom but very exact radial distance is first determined and then only the wave function is defined. The theory uses same form of wave function for both the hydrogen as well as the quarkonium applications. All the observed energy levels are used as input parameters and radial distance and the string tension are the output parameters. This assures 100\% conformance to the observed energy levels. Probably this is the only model that directly introduces the quark spin $s=1/2$ in the formalism. All relativistic effects are included right into the foundation of the theory which makes this model the simplest. The radial distance and string tension predictions are no where comparable to models based on perturbative and lattice QCD. This is because the two theories are based on the non-relativistic and semi-relativistic wave theories. The string tension is a variable parameter in this theory and a function of radial distance. Both string tension and radial distance are simultaneously established by balancing the radial distance for two spin states with common $l$. Since radial distance does not have a statistical interpretation, it should be possible to experimentally determine this parameter within the limits imposed by the uncertainty principle to futher constrain the present model. 

\section{Acknowledgment}
Author is grateful to several experts in the field for comments and suggestions.

\bibliographystyle{amsalpha}

\end{document}